\documentclass[aps,prd,reprint,showpage,showpacs,nobibnotes,superscriptaddress,twocolumn,amssymb,amsmath,nofootinbib,floatfix]{revtex4-1}
\usepackage{graphicx}
\usepackage[usenames,dvipsnames]{xcolor}
\usepackage{epsfig}
\usepackage{verbatim}
\usepackage{amsthm}
\usepackage{amsfonts}
\usepackage{amscd}
\usepackage{bm}
\usepackage[caption=false]{subfig}
\usepackage{subfig}

\begin{document}
\title{Question for $\mathsf{SU(5)}\times \mathsf{SU(5)}$ string unification}
\author{Edison T. Franco}%
\email{edisonfranco@uft.edu.br}
\affiliation{
Universidade Federal do Tocantins, Campus Universit\'ario
de Aragua\'\i na, Avenida Paraguai , Cimba\\77824-838 Aragua\'\i na, Tocantins, Brazil
}

\date{\today}

\begin{abstract}
Here we ask if it is possible to have string unification in $\mathsf{SU(5)\times SU(5)}$ gauge group. We specifically investigate the weakly coupled heterotic string unification for the four couplings in this framework. We show that only a limited versions of $\mathsf{SU(5)\times SU(5)}$ with adjoint representation components at intermediated scales, between $M_Z$ and $\Lambda$ (unification), are allowed. This is essentially due to the limitation in the parameter space to the gauge coupling constant $\alpha_1^{-1}$ related to hypercharge. Indeed, only the vanishing hypercharge decompositions of $\mathsf{SU(5)_L}$ subgroup can help to this unification, namely, the fermion and boson triplets, $\Sigma_3\sim(1,3)_0$, and the fermion and boson octets, $\Sigma_8\sim(8,1)_0$. Thus, these intermediate particles are compatible with the so-called Adjoint $\mathsf{SU(5)}$. The triplets must live in TeV region and could be accessible at colliders while the octets must alive in very high energy scales. We also show that the non-SUSY unification scenario requires the introduction of an additional $\mathsf{SU(2)_L}$ scalar color triplet, $\eta \sim (3,3)_{-{1}/{3}}$, at relatively low energies ($\sim 10^{5-11}$GeV), and it may induce the proton decay.
\end{abstract}


\maketitle


\section{Introduction}

\label{sec:intro} The unification idea, mainly in context of $\mathsf{SU(5)}$ gauge group
\cite{Georgi:1974sy}, is still as an interesting alternative for the physics
beyond the standard model (SM) \cite{Nath:2006ut}. Many unifiable models
have been proposed in this framework in recent decades in the attempt to explain
the incorporation of SM and the phenomenology that have been discovered beyond that. Unfortunately, the Minimal Supersymmetric Standard Model (MSSM)~\cite{Dimopoulos:1981zb,Sakai:1981gr}, where the best fit for the running of gauge couplings was obtained, predicts an unstable proton, which was not observed by SuperKamiokande bound~\cite{Goto:1998qg,Hayato:1999az,Murayama:2001ur}. On the other hand, a few works have shown that the appropriate Yukawa choices can induces a proton consistent with this stability bound~\cite{Bajc:2002bv,Bajc:2002pg,EmmanuelCosta:2003pu}.

While electroweak and strong interactions have their couplings easily unified around $10^{16}$GeV~\cite{Langacker:1980js,Raby:2008gh}, the quest on the unification theory of four fundamental couplings is still alive. The problem of how the quantum theory of all known interactions are accommodated into a general theory should be a task of the superstring models~\cite{EmmanuelCosta:2005nh}.

As the first requisite for any Grand Unified Theory (GUT) or Supersymmetric GUT (SUSY GUT), superstring theory must restore the SM
in its low-energy effective theory~\cite{Kakushadze:1997ne}. GUTs and SUSY GUTs are particularly favored
in the wide range of theories since they are truly unified theories with one effective gauge coupling at low-energy scale~\cite{Lewellen:1989qe,Schwartz:1989gh,Chaudhuri:1994cd,Cleaver:1995ne,Aldazabal:1994zk,Erler:1996zs,Kakushadze:1996hj}. Indeed, it has been shown that low-energy physics can be embedded in the heterotic string scheme~\cite{Gross:1984dd} and it exhibits many properties present in our range of energies~\cite{Gross:1985fr,Gross:1985rr}. This can help to solve the strong CP problem by a natural introduction of axions~\cite{Banks:1996ss}, as well as it gives a solution to the doublet-triplet-splitting problem~\cite{Pokorski:1997mv}.

Some specific group choices are very attractive to incorporate strings~\cite{Trapletti:2005ey}. In this sense, the question of what is the simplest GUT model changes to what is the most natural pattern to electroweak scale with weakly heterotic string unifications based on $E_{8}\times E_{8}$ gauge group~\cite{Maslikov:1996gn}. One possibility is the $\mathsf{SU(5)\times SU(5)}$ gauge group~\cite{Davidson:1987mi}. This symmetric theory has many attractive features, e.g., the generalized seesaw for all fermions without the need of singlets~\cite{Davidson:1987mh,Davidson:1987tr,Davidson:1989bx}. This group allows a solution to doublet-triplet-splitting problem if one considers an appropriated discrete symmetry in a GUT~\cite{Dine:2002se} or in the string scenarios~\cite{Barr:1996kp}.
Generical implications of $\mathsf{SU(5)}$-based gauge group in the LHC analysis have been studied by many works~\cite{Li:2004cj,FileviezPerez:2008ib,Dorsner:2009cu,Dorsner:2009mq,Dorsner:2010cu} and can be extended to the $\mathsf{SU(5)\times SU(5)}$ theory in a natural way.

The string unification at tree level of the three gauge couplings and the gravitational coupling is given by the following relation~\cite{Ginsparg:1987ee}
\begin{equation}
\alpha _{\mathrm{st}}=\frac{2G_{N}}{\alpha ^{\prime }}=k_{i}\alpha _{i},
\label{a_string}
\end{equation}%
where $\alpha _{\mathrm{st}}=g_{\mathrm{st}}^{2}/4\pi $ is the
string-scale unification coupling constant, $G_{N}$ is the Newton constant, $%
\alpha ^{\prime }$ is the Regge slope, $\alpha _{i}=g_{i}^{2}/4\pi $ are the gauge coupling constants, with $i$ running over the gauge groups, $\mathsf{U(1)_Y}$, $\mathsf{SU(2)_L}$, $\mathsf{SU(3)_C}$ ($i=y,w,s$\footnote{In this notation $y$ stands for h\emph{y}percharge, while $w$ stands for \emph{w}eak and $s$ stands for \emph{s}trong}), respectively. The Ka\v{c}-Moody levels in four-dimensional string depends on the unification group and make the correct adjustment to the unification. Each term in Eq.(\ref{a_string}) should be renamed as a new coupling, $k_{y}\alpha _{y}=\alpha _{1}$, $k_{w}\alpha _{w}=\alpha _{2}$, $k_{s}\alpha _{s}=\alpha _{3}$, which are indeed unifiable at $\Lambda$ scale. For the canonical groups, e.g., $\mathsf{SU(5)}$, $\mathsf{SO(10)}$, $\mathsf{E_6}$, $\mathsf{[SU(3)]^3\times Z_3}$, $\mathsf{SO(18)}$, $%
\mathsf{E_8}$, $\mathsf{SU(15)}$, $\mathsf{SU(16)}$ and $\mathsf{SU(8)\times SU(8)}$~\cite{PerezLorenzana:1997ef,PerezLorenzana:1998rj,PerezLorenzana:1999tf}, these levels are given by $%
k_{y}=5/3$, $k_{w}=1$ and $k_{s}=1$. They are different for other group choices, such
as $\mathsf{SU(5)\times SU(5)}$, $\mathsf{[SU(6)]^3\times Z_3}$, Pati-Salam models, etc. Indeed, these levels have the power to constrain the Renormalization Group Equations (RGEs) for a new class of allowed particles~\cite{EmmanuelCosta:2005nh}. In
this sense, the string nature of the unification in each theory is due to
the new equation that relates the gauge coupling constants with string scale, $\Lambda$,
\begin{equation}
\alpha _{\mathrm{st}}=\frac{1}{4\pi }\left( \frac{\Lambda }{\Lambda _{s}}%
\right) ^{2},  \label{a_string2}
\end{equation}%
and gives an additional constraint on the RGEs, where the scale $\Lambda _{s}$ is given by
\begin{equation}
\Lambda _{s}=\frac{e^{(1-\gamma )/2}3^{-3/4}}{4\pi }M_{P}\approx 5.27\times
10^{17}\mathrm{GeV},
\end{equation}
with $\gamma \approx 0.577$ being the Euler constant. Notice that the Eq.~(\ref{a_string2}) only makes sense at the unification scale for the usual gauge couplings. Thus, instead of two independent parameters in GUT, $\Lambda$ and $\alpha_U$, in string theory there is only
one parameter which relates $\Lambda$ and $\alpha_{\mathrm{st}}$.

In this paper we study this parametrization despite of the introduction of complete scalar sector. The parameter space is set univocally by low-energy multiplets in the non-SUSY and in the SUSY schemes, by the unification of SM and MSSM, respectively. This paper has
been organized as follows: in Sec.~\ref{sec:cano} we introduce the canonical
unification and the new analysis that is suitable for the string
unification; in Sec.~\ref{sec:su5su5} the general aspects of the $\mathsf{SU(5)_{L}\times SU(5)_{R}}$ theory are
introduced; in Sec.~\ref{sec:problems} we address the
problems that concerns to the coupling constants at one-loop and two-loop levels; in Sec.~\ref{sec:adj} we discuss the
inclusion of the adjoint $\mathsf{SU(5)_{L}}$ subgroup in non-SUSY
and in SUSY theories and show some examples; in Sec.~\ref{sec:pheno} phenomenological implications are
discussed; finally the summary and conclusions are the subjects of Sec.~\ref%
{sec:conclusions}.

\section{Canonical unification}

\label{sec:cano}

\subsection{Standard Model}

The renormalizability of the theory is guaranteed by a mass-dimensional scale parameter, $\mu$, which is a response of the Green functions of the theory in each energy scale. When we perform a change in $\mu$ we induces a new adjustment in the coupling constants, mass and vertex renormalization of the theory. These quantities are governed by the RGEs. The coefficients in these equations are associated with the finite shift generated by a increase in this scale parameter for each of this three constants~\cite{Machacek:1983fi,Machacek:1983tz,Machacek:1984zw}. The evolution of the gauge couplings constants at one-loop level are governed by the following RGEs:
\begin{equation}
\alpha _{i}^{-1}(\mu )=\alpha _{iZ}^{-1}-\frac{b_{i}^{z}}{2\pi }\log \frac{%
\mu }{M_{Z}},  \label{a_SM}
\end{equation}%
where $\alpha _{iZ}^{-1}=\alpha _{i}^{-1}(M_{Z})$, and we are considering the central values of electroweak constants \cite{pdg},\
\begin{eqnarray}
\alpha ^{-1} &=&127.916\pm 0.015, \\
\sin ^{2}\theta _{W} &=&0.23129\pm 0.00005, \\
\alpha _{s} &=&0.1181\pm 0.0011.
\end{eqnarray}%

If we let the Eq.(\ref{a_SM}) runs free to some high scale $\Lambda$, we do not reach unification in general (that means by Eq. (\ref{a_string}))
\begin{equation}
\left( \alpha _{y}k_{y}\right) ^{-1}\neq \left( \alpha _{w}k_{w}\right)
^{-1}\neq \left( \alpha _{s}k_{s}\right) ^{-1}
\end{equation}%

On the other hand, if we consider $N$ new thresholds, e.g., with fermions, scalars, exotic fermions, included at intermediate scales and leading to unification at $\Lambda$, Eq.(\ref{a_SM}) gives
\begin{eqnarray}
\alpha _{i-\mathrm{new}}^{-1}(\Lambda ) &=&\alpha _{i}^{-1}(\Lambda )-\frac{%
\Delta b_{i}^{1}}{2\pi }\log \frac{\Lambda }{M_{1}}-\frac{\Delta b_{i}^{2}}{%
2\pi }\log \frac{\Lambda }{M_{2}}\cdots \nonumber
\\&-&\frac{\Delta b_{i}^{N}}{2\pi }\log
\frac{\Lambda }{M_{N}}\nonumber \\
&=&\alpha _{iZ}^{-1}-\frac{b_{i}^{z}}{2\pi }\log \frac{\Lambda }{M_{Z}}-%
\frac{\Delta b_{i}^{1}}{2\pi }\log \frac{\Lambda }{M_{1}}\nonumber \\
&-&\frac{\Delta
b_{i}^{2}}{2\pi }\log \frac{\Lambda }{M_{2}}\cdots -\frac{\Delta b_{i}^{N}}{%
2\pi }\log \frac{\Lambda }{M_{N}},  \label{nthres}
\end{eqnarray}%
and, at one-loop level, the string unification of gauge couplings read as
\begin{equation}
\left( \alpha _{y-\mathrm{new}}k_{y}\right) ^{-1}= \left( \alpha _{w-\mathrm{new}}k_{w}\right)
^{-1}= \left( \alpha _{s-\mathrm{new}}k_{s}\right) ^{-1}= \alpha _{\mathrm{st}} ^{-1}.
\label{a_new}
\end{equation}%

Furthermore, at one-loop level all $\Delta b_{i}$'s are positive unless we couple new extra gauge bosons~\cite{Jones:1981we}. To avoid the fast proton decay, we will not consider gauge bosons at intermediate scales. If we consider the contribution of the second and subsequent terms at left-side of Eq.(\ref{nthres}) into an unique positive contribution, which we will denote by $f_i$ for each coupling, the number found is the necessary quantity to reach unification as follows from Eq.(\ref{nthres}). This procedure allows to choose which particle we need to include for a determined unification scale. Using this idea with Eqs. (\ref{a_SM}) and (\ref{a_new}) we rewrite Eq.(\ref{nthres}) as

\begin{eqnarray}
k_i\alpha _{\mathrm{st}} ^{-1}&=&\alpha _{i}^{-1}(\Lambda )-f_{i}(\Lambda \nonumber)\\
&=&\alpha _{iZ}^{-1}-\frac{b_{1i}^{z}}{2\pi }\log
\frac{\Lambda }{M_{Z}}-f_{i}(\Lambda ).  \label{a_string3}
\end{eqnarray}

In the above equations (there are three) we have freedom only to determine the unification scale, $\Lambda$, and the three $f_i$ functions. Notice that since we are not specifying the particles between $M_Z$ and $\Lambda$, the freedom of new particles at intermediate scales are inside $f_i$, and we have the freedom in the latter, and not only in $\Lambda$, as one would think. The string scale is constrained by $\alpha _{\mathrm{st}}$ given by Eq. $\left( \ref{a_string2}%
\right) $ which is transferred to $f_i$ functions. Then, a simplification could be obtained choosing the following parametrization
\begin{equation}
\Lambda =z\Lambda _{s},
\end{equation}%
where the parameter $z$ is contained into the
interval $1\geq z>0$ for perturbative string theory. The parameter z gives the measurement of string unification and it is only identified as the coupling $g_{\mathrm{st}}$ when all the three coupling constants meet together at the unification scale, $\Lambda=g_{\mathrm{st}}\Lambda_s$~\cite{Kaplunovsky:1992vs}. This constraint into Eq.(\ref{a_string2}) leads $\alpha _{\mathrm{st}}={z^2}/{4\pi }$. Now the Eq.(\ref{a_string3}) read as

\begin{equation}
\frac{4\pi k_{i}}{z^{2}}=\alpha _{i}^{-1}(z)-f_{i}(z),  \label{a_string3b}
\end{equation}

\begin{figure}[ptb]
\begin{centering}
\includegraphics[width=8cm]{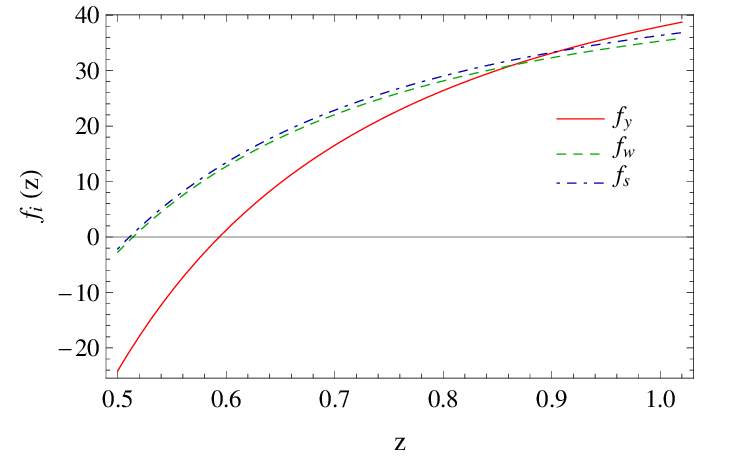}
\par
\caption{\label{fig1}Running of $f_i$ functions in terms of $z$ at one-loop level from SM at $M_Z$ to$\mathsf{SU(5)}$ string
unification.}
\end{centering}
\end{figure}

In order to unify at the perfect string unification it is necessary $z=1$ and,
therefore, it occurs when $\Lambda =\Lambda _{s}$ and with $\alpha _{\mathrm{st}%
}^{-1}=4\pi $. As $z$ decreases, the string scale also decreases while $\alpha _{\mathrm{st}}^{-1}$ increases very fast from the lower bound at $4\pi$. Using the canonical levels
and the RGE coefficients in the Appendix~\ref{sec:appendix}, the functions $f_i$ needed are showed in Fig.~\ref{fig1}. Below of $z\sim0.6$ there is no particle content which satisfies the negative values of $f_y$, thus only above zero we can specify particle content to reach string unification at $z\Lambda_s$. As near as we are of $\Lambda_s$ greater are the contribution to each coupling constant, that is, more contributions from beta coefficients are necessary. Thus, we always can look to the needed of $f_i$ given by Eqs. (\ref{a_string3b}) to determine appropriately the particle content at intermediate scales.

Notice that, in Fig.~\ref{fig1} the necessary $f_y$, $f_w$ and $f_s$ have the same order for $ z \gtrsim 0.6$ although their very different symmetry origin. For a zoom in the values of $z \gtrsim 0.9$ these functions shows that all of them are in the range $32\lesssim f_i \lesssim 38$ at one loop level.

\begin{figure}[ptb]
\begin{centering}
\includegraphics[width=8cm]{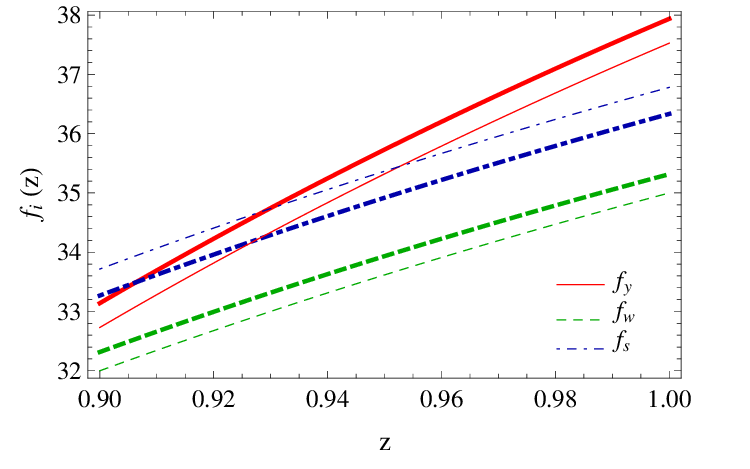}
\caption{Running of $f_i$ functions in terms of $z$ at one-loop level (thick lines) and two-loop level (thin lines) from SM at $M_Z$ to $\mathsf{SU(5)}$ string unification.}
\label{fig2}
\end{centering}
\end{figure}

The two-loop level analysis should provide a more precise values of $f_i$ functions~\cite{Machacek:1983fi,Machacek:1983tz,Machacek:1984zw,Barger:1992ac}. At this level, although we cannot find analytical solutions to the functions $f_{i} $, we can solve numerically the RGEs for couplings (without inclusion of intermediate scales) and then evaluate the values of that functions just by a imposed unifying string point. Solving these equations at electroweak scale and running it to unification scale, $\Lambda$, we must solve the following set of differential equations for the gauge couplings
\begin{equation}
\frac{\partial \alpha _{i}^{-1}(t)}{\partial t}+\frac{1}{2\pi }\left(
b_{i}^{z}+\frac{1}{4\pi }b_{ij}^{z}\alpha _{j}(t)\right) =0,  \label{RGE_2}
\end{equation}%
where $b_{ij}^z$ are the $\beta$-function coefficients of RGE in the Appendix~\ref{sec:appendix}. After solving that we use Eqs.(\ref{a_string3b}). Fig.~\ref{fig2} shows $f_i$ for $z\gtrsim 0.9$ and shows that the string unification is naturally reached in this context since there is no preferential particle content,  at least in this restricted range in the canonical unification.

Besides the string theory could be
implemented in such non-SUSY scheme~\cite{Dienes:1996du} this is rather unnatural
since superstring theory includes natively the supersymmetry~\cite{Chkareuli:1998wi,Chkareuli:1998vx}. In this sense, let us introduce the SUSY scenario.

\subsection{Minimal Supersymmetric Standard Model}

\begin{figure}[ptb]
\begin{centering}
\includegraphics[width=8cm]{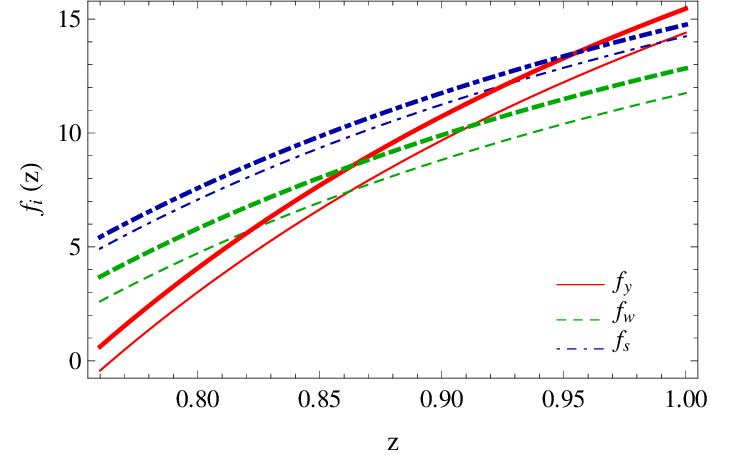}
\caption{Running of $f_i$ functions in terms of $z$ at one-loop level (thick lines) and two-loop level (thin lines) from SM at $M_Z$ to MSSM at $1$TeV and then to$\mathsf{SU(5)}$ string unification.}
\label{fig3}
\end{centering}
\end{figure}

The supersymmetrization of SM is a well-motivated model extension and it could
be tested in the collision experiments, for example, at LHC~\cite{Hinchliffe:1996iu,ArkaniHamed:2004fb}. The minimal version,
the MSSM, changes SM RGE coefficients at the SUSY scale $M_S$. Using the beta coefficients at Appendix~\ref{sec:appendix}, with $M_S=1$ TeV, the SUSY (in DR-scheme) $f_{i}$ functions at one- and two-loop order are showed in Fig.~\ref{fig3} for $z\gtrsim 0.75$. Here we can see that the $f_{y}$ is the most restrictive function since it is only positive in a very small range of $z$. In this sense, this result shows why SUSY string unification is easier to reach than non-SUSY case for canonical groups (compare with Fig.~\ref{fig2}). The introduction of only few particles can reach the string unification. We have tested and there is no considerable variation even if SUSY scale is shifted from 1 TeV to $M_{Z}$. We will see in the next section that it is not too easy to unify into a group with non-canonical Ka\v{c}-Moody levels, as is the case of the $\mathsf{SU(5)_{L}\times
SU(5)_{R}}$ gauge group.

\section{$\mathsf{SU(5)_{L}\times SU(5)_{R}}$ Model}

\label{sec:su5su5}
Let us extend the analysis of previous section to a non-canonical $\mathsf{SU(5)\times SU(5)}$ gauge theory changing the Ka\v{c}-Moody levels~\cite{Dienes:1995sq,Cho:1997gm,Barger:2005qy,Barger:2005gn}. The study of weakly interacting heterotic string is also a motivation for the use of crossed gauge group since it is possible to have a path to unification into the string gauge
group $\mathsf{E_{8}\times E_{8}}$. On the other hand, this changes dramatically the number of particles which must be accommodate at intermediate scales. The gauge group $\mathsf{SU(5)_{L}\times SU(5)_{R}}$, specifically with the left-handed and right-handed sectors separated in each $\mathsf{SU(5)}$ subgroup, maintain Georgi-Glashow left-handed feature with a non-minimal modification. The non-SUSY $\mathsf{SU(5)_{L}\times SU(5)_{R}}$ has the following fermionic representations\footnote{The notation stands for $\sim$ (rep. of $SU(5)_L$, rep. of $SU(5)_R$).}
\begin{eqnarray}
\psi_{L}&\sim &(\overline{5},1)=
\left[
\begin{array}{c}
D_{1}^{c} \\
D_{2}^{c} \\
D_{3}^{c} \\
e \\
-\nu%
\end{array}%
\right]_{L}, \quad
\psi _{R}\sim(1,\overline{5})=\left[
\begin{array}{c}
D_{1}^{c} \\
D_{2}^{c} \\
D_{3}^{c} \\
e \\
-\nu%
\end{array}%
\right]_{R},
\end{eqnarray}
\begin{eqnarray}
\chi _{L}&\sim &(10,1)=\frac{1}{\sqrt{2%
}}\left[
\begin{array}{ccccc}
0 & U_{3}^{c} & -U_{2}^{c} & -u_{1} & -d_{1} \\
-U_{3}^{c} & 0 & U_{1}^{c} & -u_{2} & -d_{2} \\
U_{2}^{c} & -U_{1}^{c} & 0 & -u_{3} & -d_{3} \\
u_{1} & u_{2} & u_{3} & 0 & -E^{c} \\
d_{1} & d_{2} & d_{3} & E^{c} & 0%
\end{array}%
\right] _{L}
\end{eqnarray}
\begin{eqnarray}
\chi _{R} &\sim &(1,10)=\frac{1}{\sqrt{2%
}}\left[
\begin{array}{ccccc}
0 & U_{3}^{c} & -U_{2}^{c} & -u_{1} & -d_{1} \\
-U_{3}^{c} & 0 & U_{1}^{c} & -u_{2} & -d_{2} \\
U_{2}^{c} & -U_{1}^{c} & 0 & -u_{3} & -d_{3} \\
u_{1} & u_{2} & u_{3} & 0 & -E^{c} \\
d_{1} & d_{2} & d_{3} & E^{c} & 0%
\end{array}%
\right]_{R}.
\label{SU5L}
\end{eqnarray}

Notice that the introduction of SUSY does not introduce more fermion representations, but all the superfields should be written in the left-handed fashion, which could be made straight~\cite{Willenbrock:2004hu}. The conventional fermions' colors, flavors and electric charges are naturally
indicated by the notation. Here we are denoting the SM particles, and the SM neutrino counterpart, $\nu_R$, by
lowercase letters, while the capital letters indicates the three vector-like fermions, $U$, $D$ and $E$, that
are naturally introduced in a left-right theory to fulfill the multiplets.

In this GUT we have the freedom to choose some paths to the symmetry-breaking to electromagnetic theory
due to a nontrivial potential that could be introduced. We show bellow explicitly two suitable symmetry-breaking patterns which accommodates our proposal~\cite{Cho:1993jb}
\begin{equation*}
\begin{array}{c}
SU(5)_{L}\times SU(5)_{R} \\
\downarrow \Lambda \\
SU(3)_{L}\times SU(2)_{L}\times U(1)_{L}\times SU(3)_{R}\times
SU(2)_{R}\times U(1)_{R} \\
\downarrow M_{LR} \\
SU(3)_{L+R}\times SU(2)_{L}\times SU(2)_{R}\times U(1)_{L+R} \\
\downarrow M_{R} \\
SU(3)_{L+R}\times SU(2)_{L}\times U(1)_{Y} \\
\downarrow M_{W} \\
SU(3)_{L+R}\times U(1)_{EM},%
\end{array}%
\end{equation*}%
where we identify $\mathsf{SU(3)_{L+R}=SU(3)_{C}}$, and both $\mathsf{SU(5)}$ groups are
being broken simultaneously at $\Lambda$, and
\begin{equation*}
\begin{array}{c}
SU(5)_{L}\times SU(5)_{R} \\
\downarrow \Lambda_L \\
SU(3)_{L}\times SU(2)_{L}\times U(1)_{L}\times SU(5)_{R} \\
\downarrow \Lambda_{R} \\
SU(3)_{L+R}\times SU(2)_{L}\times U(1)_{Y} \\
\downarrow M_{W} \\
SU(3)_{C}\times U(1)_{EM},%
\end{array}%
\end{equation*}%
with the $\mathsf{SU(5)_L}$ being broken at $\Lambda_L$, higher than the breaking of $\mathsf{SU(5)_R}$ at $\Lambda_R$. Neither of the patters must be imposed in our analysis since, in approximation, we can choose the simultaneous breaking of all intermediate symmetries at $\Lambda^{(\mathrm{all})}$ as
\begin{equation*}
\begin{array}{c}
SU(5)_{L}\times SU(5)_{R} \\
\downarrow \Lambda^{(\mathrm{all})} \\
SU(3)_{L+R}\times SU(2)_{L}\times U(1)_{Y} \\
\downarrow M_{W} \\
SU(3)_{C}\times U(1)_{EM}.%
\end{array}%
\end{equation*}

If we use $f_i$ functions no
symmetry breaking analysis are necessary at the present discussion, but a further analysis would be required to determine the full setup, which is beyond the scope of this paper. Let us run the RGEs starting only with one Higgs scalar at low energies in the non-SUSY unification. In the SUSY unification we will consider the two MSSM scalars at scale $M_S$ scale as well as all their superpartners.

Mathematically, if $%
\alpha_i$ is the coupling constant of invariant subgroup $G_i$, the
numerical factors read as~\cite{PerezLorenzana:1998rj}
\begin{equation}\label{k_i}
\frac{1}{k_i}\equiv \frac{\alpha_i}{\alpha}=\frac{\mathrm{Tr} T^2}{\mathrm{Tr%
} T_i^2},
\end{equation}%
where $T$ is a generator of the SM subgroup $G_i$, properly normalized over a
representation $R$ of $G$, and $T_i$ is the same generator but normalized
over the representation of $G_i$ embedded in $R$, with the traces running
over complete representations. Thus, the only possible values for $k_i$ ($i=w,s$)
are integer numbers~\cite{PerezLorenzana:1997ef,PerezLorenzana:1998rj,PerezLorenzana:1999tf}. There are two $\mathsf{SU(3)}$ from each $\mathsf{SU(5)}$ while the $\mathsf{SU(2)_L}$ comes from $%
\mathsf{SU(5)_{L}}$, thus $k_s=2$ and $k_w=1$. The hypercharge factor cannot be immediately inferred
since it depends on the combinations inducing to the $U(1)_{Y}$ in the
branch. This is $k_y=13/3$ in the present group.

The $\sin ^{2}\theta _{W}$ depends on the type of unification of
different gauge couplings for each of $\mathsf{SU(5)}$ invariant subgroup. For a perfect left-right symmetry one can show that at unification scale
\begin{equation}
\sin ^{2}\theta _{W}(\Lambda )=\frac{1}{1+k_{y}/k_{w}}=\frac{3}{16}.
\end{equation}%

For canonical groups $\sin ^{2}\theta _{W}$ increases with scale and, at unification scale, it is always bigger than the value at electroweak scale. Indeed it were believed that $\sin ^{2}\theta$ should increase as the energy increases regardless of what would be introduced at intermediate scale~\cite{Mohapatra:1996fu,Mohapatra:1996iy}. However, we have shown that this is not always true since the introduction of $(15+\overline{15},1)+(1,15+\overline{15})$ play the main role to reconcile the values at $\Lambda$ and at electroweak scale~\cite{EmmanuelCosta:2011wp}. It opens the range of possibilities to unification and allows to reached $\sin ^{2}\theta _{W}(\Lambda )=3/16$ consistently once the content at intermediate scale is properly selected.

The right-handed neutrino is natural and there is no singlets in this theory. Once the suitable scalar representation is chosen, the
neutrino can have their mass naturally generated. For instance, it is distinguishable from mirror symmetry~\cite{Collie:1998ty}, since the introduction of $(\overline{5},5)$ scalar representation allows a natural Dirac mass term to neutrino. The introduction of $(1,15)$ scalar
representation does not mix the right-handed with left-handed sector, thus the neutrinos get only a Majorana mass term. The difference between the above theory to the usual Georgi-Glashow theory is that we need an crossing scalar to generate the most simple Dirac mass term for any fermion once gauge singlets are not included. The universal seesaw mechanism, which
naturally works for all charged leptons and vector-like fermions~\cite{Davidson:1987tr,Cho:1993jb},
can also works for neutrino sector~\cite{Davidson:1987mh}. Once for neutrinos there is
no vector-like fermion associated, the seesaw mechanism works like a
generalized type II seesaw mechanism~\cite{Magg:1980ut,Schechter:1980gr,Mohapatra:1980yp}. The study of seesaw mechanism for $\mathsf{SU(5)\times SU(5)}$ deserves a further analysis and will be considered elsewhere.

\section{$\mathsf{SU(5)\times SU(5)}$ string unification}
\label{sec:problems}
\subsection{non-SUSY $\mathsf{SU(5)\times SU(5)}$}

\begin{figure}[ptb]
\begin{centering}
\includegraphics[width=8cm]{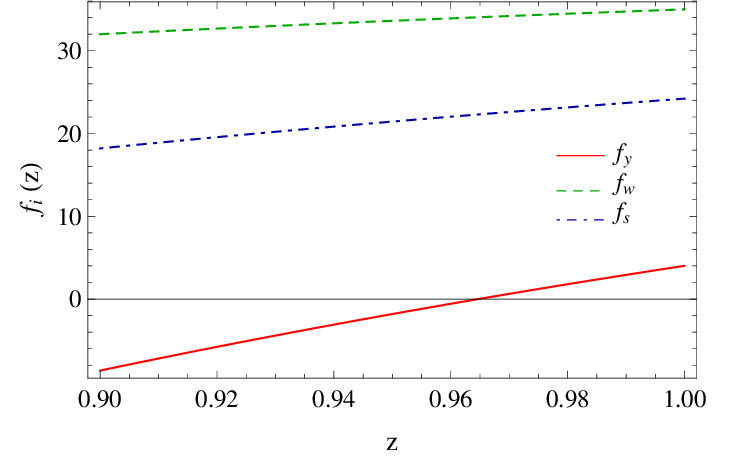}
\caption{Running of $f_i$ functions in terms of $z$ at two-loop level from SM at $M_Z$ to $\mathsf{SU(5)\times SU(5)}$ string unification.}
\label{fig4}
\end{centering}
\end{figure}

Let us consider the Eqs.(\ref{a_string}) and (\ref{RGE_2}) at two-loop level with SM content at $M_{Z}$ scale. The difference from the canonical analysis in Sec. \ref{sec:cano} is the introduction of the new non-canonical levels. We show the $f_i$ functions needed to string unification in Fig.\ref{fig4}. There we can notice that the theory only reaches unification with the parameter of the order of $z\gtrsim0.96$. Thus, this scenario is unfavorable since that are needed a very large contribution to
$f_{w} \simeq 33$ and to $f_{s}\simeq 22$, while a very small
contribution to $f_{y}$ must be introduced. From this view, the function $f_{y}$ is
strongly constrained in the range $0.96\gtrsim z\gtrsim 1.00$. This is a consequence of $U(1)$
decomposition inside the group structure, and its abundant contribution leads to the level
$k_{y}=13/3$, greater than the canonical value, $5/3$. All particles induces contributions which are absorbed into $f_{i}$ functions as a parameter redefinition. It means that new particle content needs to be introduced but with only contributions with small hypercharge or indeed with vanishing hypercharge. On the other hand, almost all particles have non-vanishing hypercharge (this is the most abundant quantum number in the theory). Consequently it is very difficult to reach the string unification in this case. In Sec.\ref{sec:adj} we show a solution in the context of Adjoint $\mathsf{SU(5)}$ with a three fermion in the adjoint representation at some specific intermediate scale.

\subsection{SUSY $\mathsf{SU(5)\times SU(5)}$}

\label{subsec:SUSYsu5su5}

\begin{figure}[ptb]
\begin{centering}
\includegraphics[width=8cm]{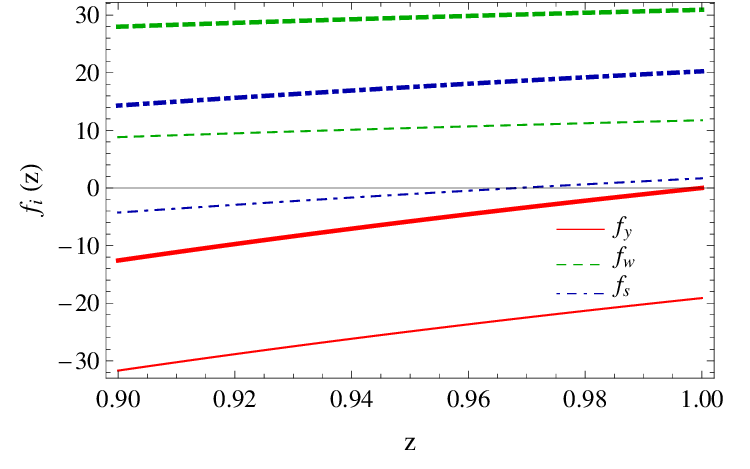}
\caption{Running of $f_i$ functions in terms of $z$ at two-loop level in $\mathsf{SU(5)\times SU(5)}$ unification. The scale of MSSM is at $M_{S}=1$TeV (thin lines) and at $M_{S}$ $\simeq 1.3\times 10^{15}$GeV (thick lines). In the latter case we get $f_y(z=1)=0$.}
\label{fig5}
\end{centering}
\end{figure}

Let us considere the SUSY case, with $M_S$ at 1TeV. In this case we also have the non-canonical Ka\v{c}-Moody levels. The Fig.~\ref{fig5} shows that we cannot obtain string unification since the function $f_{y}$ is always negative at two-loop level. Furthermore, the function $f_{s}$ is only positive for $z \gtrsim 0.97$. Notice that even with introduction of some gauge bosons at intermediate scales the function $f_{y}$ can not be fitted since the gauge group $U(1)_{Y}$ does not
contribute to negative $\beta $-coefficients (looking to one-loop level RGE beta coefficients). In this sense the study of the minimal energy for $M_S$ scale is necessary.

If we choose the scale of $f_{y}$ to matches to perfect unification at $z=1$, allowing the $M_S$ to high values, we obtain
$M_S \simeq 1.3\times 10^{15}$ GeV at two-loop level. In this case there is no more contribution needed to $f_{y}$ function. This is also showed in the Fig.~\ref{fig5} (thick lines).

The procedure employed above allows to replace the SUSY scale exactly to set $f_{y}=0$, and thus there is no need of hypercharge contribution to get perfect string unification. Now we need to introduce some new content with
vanishing hypercharge with enough contribution to $f_{w}\simeq 20$ and $f_{s}\simeq 30$, as we can see from Figs.~\ref{fig4} and ~\ref{fig5}. These contributions are only possible if we introduce scalars and fermions in Adjoint representation below $M_S$ scale or in the non-SUSY case. The fermions in Adjoint representations are necessary since purely scalar sector at very low-energy does not induces sufficient contribution even at $M_Z$ scale. We discuss it in details in the next section.

\section{Unification with Adjoint $\mathsf{SU(5)}_L$ representations}
\label{sec:adj}

\subsection{Model}

The study of adjoint fermionic content was proposed to correct the unification by the inclusion of light fermionic triplets and it is known that it correctly accommodates the neutrino mass with a mixed type I+III seesaw mechanism~\cite{Bajc:2006ia}.
Indeed, many works and several applications have been considered in Adjoint $\mathsf{SU(5)_{L}}$~\cite{Dorsner:2006fx,Perez:2007rm}, as the seesaw mechanism to neutrinos~\cite{Bajc:2007zf} or the use of family
symmetry with SUSY~\cite{Cooper:2010ik}. In this section we will briefly
introduce the string unification using this setup.

In our study there are need three adjoint fermion generations to allow the correct values of the $f_{w}$ and $f_{s}$ functions. The introduction of triplets and octets of $(24,1)$ representation in the intermediate scales fits the couplings to the correct unification scale\footnote{The decomposition to SM representation is $(\mathbf{24},1)=\left( \mathbf{\left( 1,3\right) +\left( 8,1\right) +\left( 3,2\right)
+\left( \bar{3},2\right) +\left( 1,1\right)} ,1\right)$, where the notation of the subdecomposition of $\mathsf{SU(5)_L}$, is (rep. of $SU(3)$, rep. of $SU(2)_L$)}. Let us consider inside the $\mathsf{SU(5)_L}$ just $N_{\Sigma_3}^F$ fermionic triplets multiplets, $\Sigma_3^F\sim((1,3)_0,1)$, and $N_{\Sigma_8}^F$ fermion octets, $\Sigma_8^F\sim((8,1)_0,1)$.
We can also use the regular scalar $(\mathbf{24},1)$ representation in similar way, with $N_{\Sigma_3}^S$ scalar triplets and $N_{\Sigma_8}^S$ scalar octets. Lowering the scale to these particles with vanishing hypercharge at an intermediate step does not spoil the unification, e.g, by an unstable proton, but rather there are a few works in compatibility with experimental proton lifetime which are described in the framework of Adjoint $\mathsf{SU(5)}$ at
TeV scale~\cite{Bajc:2007zf}. Yet the neutral component of the fermion triplet $\Sigma _{3}^F$ could be a
candidate to the Cold Dark Matter in this case~\cite{Perez:2008ry}.

Let us suppose that any of the right-handed multiplets are at unification scale. In order to induce adequate $f_{w}$ and $f_{s}$
contributions, in the non-SUSY as well as in SUSY case, we need three (generations of) multiplets of each adjoint kind ($N_{\Sigma_3}^F$=$N_{\Sigma_8}^F$=3). As these multiplets have vanishing hypercharge, at one loop level the $f_{y}$ does not suffer any modification, but at two-loop level a shift is obtained due to the interplay with other two gauge couplings. Let us consider the non-SUSY and the SUSY cases separately.

\subsection{Non-SUSY String Unification}

The non-supersymmetric case is more restrictive than high-energy SUSY case. Even if fermion and scalar $\Sigma _{3}$ are at $M_{Z}$, the function $f_{w}$ is still negative. If we use brute force to save the unification we need to introduce the scalar $\eta \sim ((3,3)_{-1/3},1)$ of $\mathsf{SU(5)_L}$. Their contribution is showed in Appendix~\ref{sec:appendix_b}. This particle is inside the representation $H=(45,1)$, and gives relevant contribution to $f_{w}$, thus it drives the gauge couplings to perfect string unification at $z=1$. There is no other particle in this model which can corrects the running of gauge couplings to string unification in non-SUSY case.

\begin{figure}
\center
\subfloat[\label{fig6a}]{\includegraphics[width=8cm]{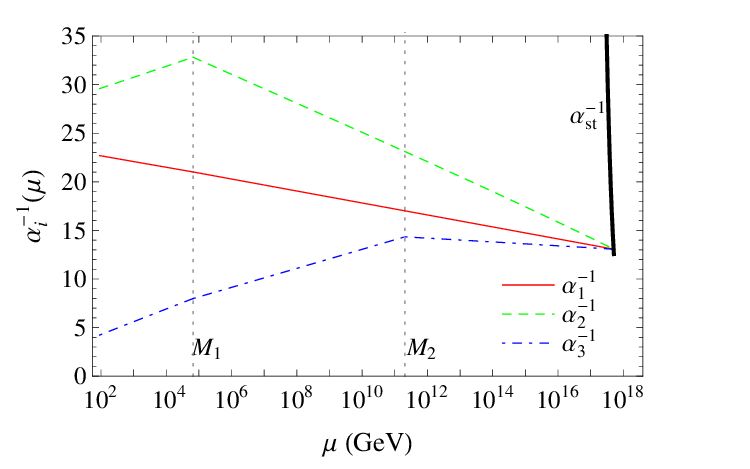}}
\hfill
\subfloat[\label{fig6b}]{\includegraphics[width=8cm]{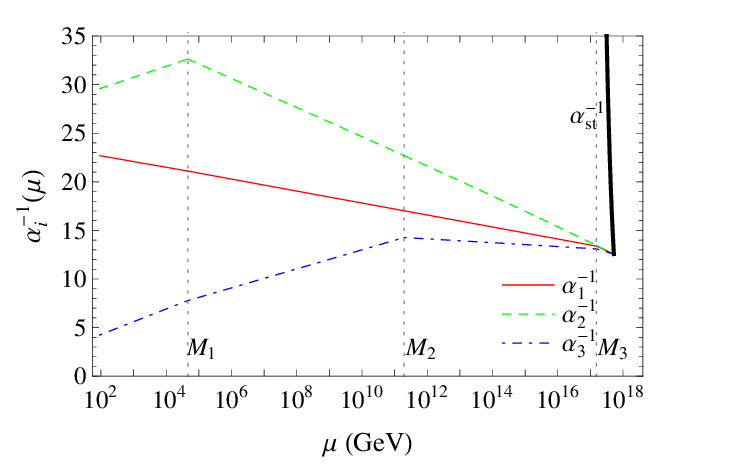}}
\hfill
\subfloat[\label{fig6c}]{\includegraphics[width=8cm]{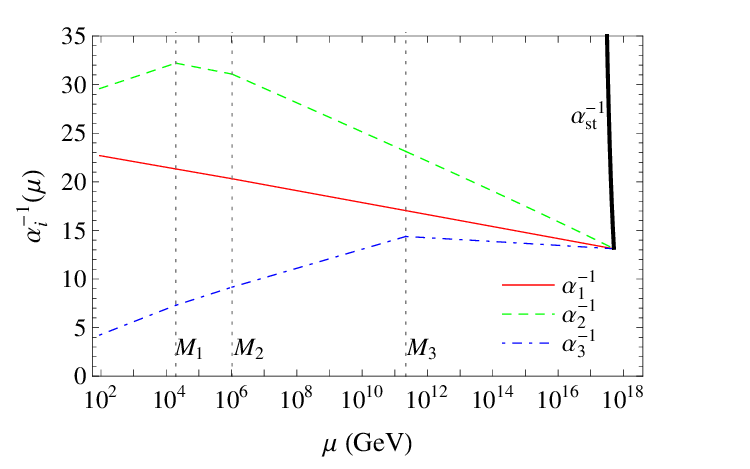}}
\caption{ Example of non-SUSY $\mathsf{SU(5)\times SU(5)}$ string unification at two-loop level. All the runnings starts from SM at $M_Z$.  In (a) at $M_1 \sim 10^5$GeV we introduce the scalars $\Sigma_3^S$,$\Sigma_8^S$, three generations of fermion $\Sigma_3^F$and the $\eta$ while at $M_2 \sim 10^{11}$GeV there are introduced three generations of $\Sigma_8^F$. In (b) is the same as (a) but with vector-like fermions at $M_3 \sim 10^{17}$GeV. In (c) at $M_1 \sim 10^4$GeV we introduce the scalars $\Sigma_3^S$, $\Sigma_8^S$ and three generations of $\Sigma_3^F$. At $M_2\sim 10^6$GeV there are introduced the $\eta$ and at $M_3 \sim 10^{11}$GeV there are introduced three generations of $\Sigma_8^F$. }
\label{fig9}
\end{figure}

The Fig.{\ref{fig6a}} shows the string unification at $\Lambda_s$ (with some freedom, as we discuss below) in $\mathsf{SU(5)\times SU(5)}$. The running starts with SM content and at  $M_1\sim 10^{4-5}$GeV there are introduced the scalars $\Sigma_3^S$, $\Sigma_8^S$ and $\eta$, and three generations of $\Sigma_3^F$. At $M_2\sim 10^{11}$GeV there are introduced three generations of $\Sigma_8^F$. The Fig.~\ref{fig6b} shows the same as Fig.{\ref{fig6a}}, but at $M_3\sim 10^{17}$GeV we have introduced the vector-like fermions to finish the unification. Notice that in the two cases above the scale of $\eta$ is very low, which can induces a fast proton decay. To circumvent the problem we must enhance the $\eta$ scale. The Fig.~{\ref{fig6c}} shows the unification at $\Lambda_s$. Now, the $\Sigma_3^S$, $\Sigma_8^S$ and three fermion generations of $\Sigma_3^F$ are included at $M_1 \sim 10^4$GeV and $\eta$ is shifted to $M_2 \sim 10^6$GeV. To complete the unification, at $M_3 \sim 10^{11}$GeV we need to introduce three generations of $\Sigma_8^F$.

\begin{figure}
\center
\subfloat[\label{fig7a}]{\includegraphics[width=8cm]{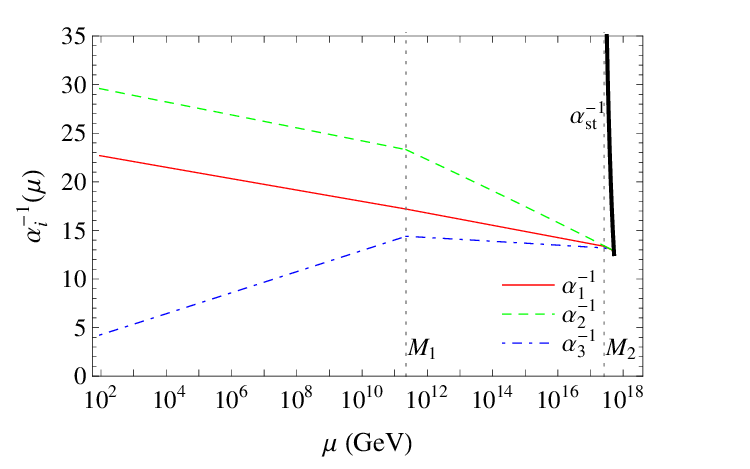}}
\hfill
\subfloat[\label{fig7b}]{\includegraphics[width=8cm]{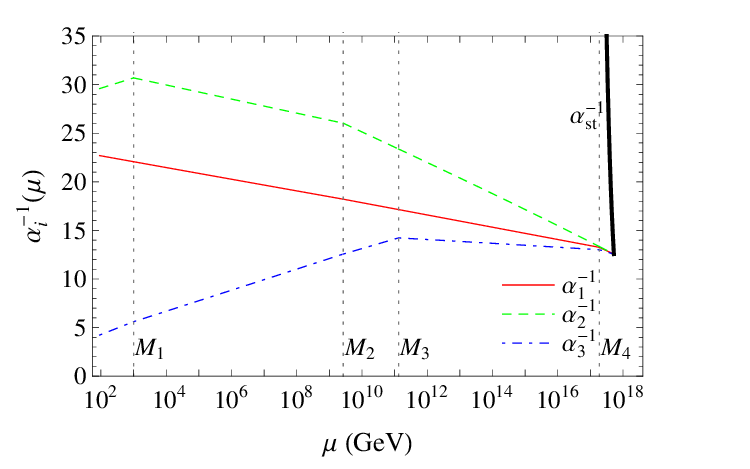}}
\caption{Example of non-SUSY $\mathsf{SU(5)\times SU(5)}$ string unification at two-loop level. All the running starts from SM at $M_Z$.  In (a) the $\Sigma_3^S$, $\Sigma_8^S$ and three generations of $\Sigma_3^F$ are also introduced at $M_z$. At $M_1 \sim 10^{11}$GeV we have introduced $\eta$, a second Higgs doublet, H, and three generations of $\Sigma_8^F$, while at $M_2 \sim 10^{17}$GeV we have to introduced the vector-like fermions to adjust the unification.
In (b) the running starts at SM with one additional Higgs doublet. The scale $M_{1} \sim 1$TeV is
where three generations of $\Sigma _{3}^F$, one scalar $\Sigma _{3}^S$ and one $\Sigma _{8}^S$ are introduced and $M_{2} \sim 10^9$GeV is where $\eta $ is included, the $M_{3} \sim 10^{11}$GeV is the scale of the three generations of $\Sigma _{8}^F$ and $M_{4}$ is the scale of vector-like fermions.}
\label{fig7a_7}
\end{figure}

The Fig.~{\ref{fig7a}} shows the extreme case when all triplets, three fermion generations of $\Sigma_3^F$ and one $\Sigma_3^S$, and one scalar octed, $\Sigma_8^S$, are at $M_z$. The three generations of $\Sigma_8^F$, $\eta$ and an additional scalar doublet, $H$, must be included at the intermediate scale at $M_1 \sim 10^{11}$GeV. The vector-like fermions are also included near to unification to do the convergence of four couplings at single point. The additional $\mathsf{SU(2)_L}$ doublet, $H$, is a priori a component of $(45,1)$\footnote{$(\mathbf{45},1)=(\mathbf{(1,2)}+\mathbf{(3,1)}+\mathbf{(3,3)}
+\mathbf{(\bar{3},1)}+\mathbf{(\bar{3},2)}+\mathbf{(\bar{6},1)}+\mathbf{(8,2)},1)$}. We have also run the RGEs starting from electroweak scale with an additional $H$ at $M_Z$. This scenario is showed in Fig.~\ref{fig7b}, where at the scale $M_{1}\sim 1$TeV three generations of $\Sigma _{3}^{F}$, an scalar triplet, $\Sigma _{3}^{S}$, and an scalar octet, $\Sigma _{8}^{S}$, are
introduced, $\protect\eta $ is shifted to $M_{2}\sim 10^9$GeV, the $M_{3}\sim 10^{11}$GeV is the
scale of three generations of fermion octets, $\Sigma _{8}^{F}$, and $M_{4}\sim 10^{17}$GeV is the scale of vector-like
fermions.

The unification examples showed above illustrates how the $\alpha_y$ is constrained. All Figs.~\ref{fig6a},~\ref{fig6b},~\ref{fig6c},~\ref{fig7a} and~\ref{fig7b} shows that $\alpha_1^{-1}$ runs directly to the string scale in some automatic form. Due to that, we are only allowing new particles without hypercharge contribution ($\Sigma$'s). Unfortunately, we cannot reach string unification to the $\alpha_2^{-1}$ automatically: it starts from $M_Z$ scale with a positive slope and we need to decrease it to negative values in order to reach for the $\alpha_1^{-1}$ coupling. To make it efficiently we introduce $\eta$, since it gives the most minimal contribution to the $\alpha_3^{-1}$ sloop, the coupling we do not want to touch. Thus, we have found solutions only with $\eta$ at relativity low scales while the vector-like must live at very-high energy, near the unification scale. These solutions are not arbitrary. Indeed, we do not have find consistent solutions with $z=1$ in the ordering of intermediate scales as showed in the Table~\ref{table1}.

\begin{table}
\caption{\label{table1} Scale ordering without unification at $z=1$ (VLF={vector-like fermions}, H={additional Higgs doublet})}
\begin{ruledtabular}
\begin{tabular}{c|cccccc}
   Order & Case 1 & Case 2 & Case 3 & Case 4 & Case 5 &  Case 6 \\ \hline \hline
  $M_Z$ & SM & SM & SM & SM & SM,$\Sigma_3^S$ & SM\\ \hline

  $M_1$ & $\Sigma_3^S$,$\Sigma_8^S$ & 3$\Sigma_3^F$ & 3$\Sigma_3^F$ & $\Sigma_3^S$,$\Sigma_8^S$ & 3$\Sigma_3^F$ & $\Sigma_3^S$,3$\Sigma_3^F$ \\ \hline

  $M_2$ & 3$\Sigma_3^F$ & $\Sigma_3^S$,$\Sigma_8^S$ & $\Sigma_3^S$,$\Sigma_8^S$ & 3$\Sigma_3^F$ & $\Sigma_8^S$ & $\Sigma_8^S$,3$\Sigma_8^F$ \\ \hline

  $M_3$ & $\eta$ & $\eta$ & $\eta$,$H$ & $\eta$,$H$ & $\eta$,$H$ & $\eta$,$H$   \\ \hline

  $M_4$ & 3$\Sigma_8^F$ & 3$\Sigma_8^F$ & 3$\Sigma_8^F$ & 3$\Sigma_8^F$ & 3$\Sigma_8^F$ & -  \\ \hline

  $M_5$ & VLF & VLF & VLF & VLF & VLF & -  \\
\end{tabular}
\end{ruledtabular}
\end{table}

\subsection{High energy SUSY}

We have seen in Sec.~\ref{subsec:SUSYsu5su5} that SUSY is only compatible when this is included at high-energy scales, very close to the string unification scale. Some previous studies in the SM context have also been focused in this way~\cite{Barger:2005qy,Barger:2005gn}. This type of SUSY scale is favored by string landscape~\cite{Bousso:2000xa,Kachru:2003aw,Susskind:2003kw,Denef:2004ze}. The inclusion of adjoint sectors (fermion and boson) bellow high $M_S$ scale\footnote{Notice that bellow the SUSY scale we only include particles in the non-SUSY way and then supersymmetrize them as the SUSY scale is reached.} allows to unify the gauge couplings correctly. Furthermore Adjoint $\mathsf{SU(5)}$ theory can also accommodate the supersymmetry~\cite{Perez:2007iw}. GUT models based on $\mathsf{SO(10)}$ symmetry and Adjoint scalar sector are also compatible with superstring models~\cite{Chaudhuri:1994ri}.

\begin{figure}[ptb]
\begin{centering}
\includegraphics[width=8cm]{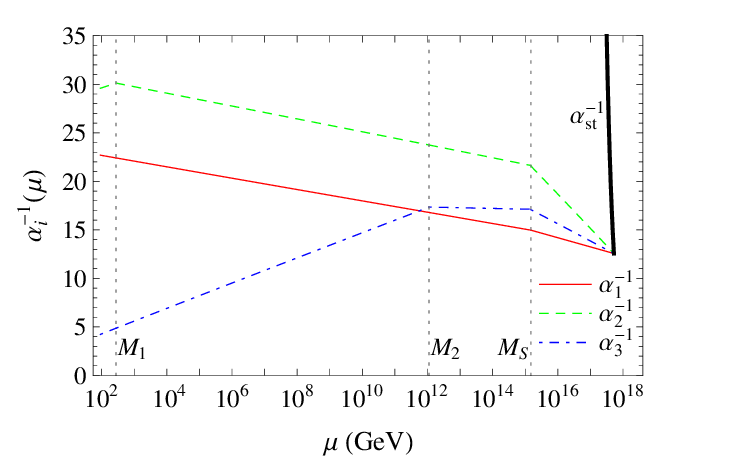}
\end{centering}
\caption{Example of SUSY $\mathsf{SU(5)\times SU(5)}$ string unification at two-loop level. Now, at $M_{1}\sim 300$GeV there are included three generation of $\Sigma _{3}^F$ and one scalar $\Sigma _{3}^{S}$ while the $M_{2}\sim 10^{12}$GeV is the scale where three generations of $\Sigma _{8}^{F}$ and one scalar $\Sigma _{8}^{S}$ are introduced. The SUSY scale is at $M_{S}\sim 10^{15}$GeV.}
\label{fig8}
\end{figure}

As an illustration we show in the Fig.~\ref{fig8} the string
unification with $z=1$. All fermions and bosons triplets, three $\Sigma_3^F$ and one $\Sigma_3^S$, are at $M_1 \sim 10^{2.5}$GeV, while we set the three $\Sigma_8^F$ and one $\Sigma_8^S$ at $M_2 \sim 10^{12}$GeV. The scale of supersymmetry is very high, $M_S \sim 10^{15}$GeV. In this case the vector-like fermions are not necessary. Comparing Fig.~\ref{fig7b} with Fig.~\ref{fig8} we can notice that the role of the SUSY in the second case is indeed equivalent to the role played by vector-like fermions in the former. In addition, this scheme shows an economical outline for SUSY string unification from the particle content point of view. The low scale of fermion triplet $\Sigma _{3}^F$ is compatible with few works where neutrinos get mass through the type I+III seesaw mechanisms for Adjoint $\mathsf{SU(5)}$~\cite{Dorsner:2006fx,Arhrib:2009mz,Awasthi:2010xc}.

\begin{figure}[ptb]
\begin{centering}
\includegraphics[width=8cm]{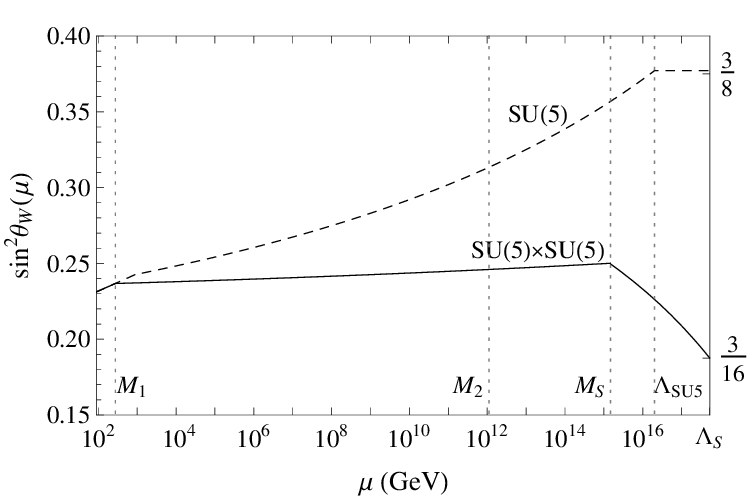}
\caption{The dashed line shows the running of $\mathsf{sin^2(\theta_w)}$ at two-loop order in the MSSM ($M_{S}=1$TeV) to unify in $\mathsf{SU(5)}$ at $\Lambda_{SU5}\simeq 2 \times 10^{16}$GeV without intermediate scales. The solid line represents the string unification in $\mathsf{SU(5)\times SU(5)}$ (the case showed in Fig.~\ref{fig8}). The scales indicated here by vertical dashed lines are only relative to the $\mathsf{SU(5)\times SU(5)}$ group. Notice also that at the scale $M_{2}\sim 10^{12}$GeV there is no changing in the $\mathsf{SU(5)\times SU(5)}$ slope since there we only include colored particles, which do not contribute to $\mathsf{sin^2(\theta_w)}$.}
\label{figsin1}
\end{centering}
\end{figure}

In order to see the consistence of the running of $\mathsf{sin^2(\theta_W)}$ in the SUSY $\mathsf{SU(5)} \times \mathsf{SU(5)}$ string theory to the correct value $3/16$ we have shown a comparison in Fig.~\ref{figsin1} with the usual $\mathsf{SU(5)}$ MSSM at two-loop level. This figure confirms that the running drives the $\mathsf{sin^2(\theta_W)}$ to a lower values in string unification based on $\mathsf{SU(5)\times SU(5)}$ group while in the canonical $\mathsf{SU(5)}$ MSSM it goes to nearly the usual $3/8$.

\subsection{Proton decay}

The non-SUSY extension can induces the proton decay through $\eta$. Indeed, due to the
introduction of $(45,1)$-scalar the proton could decay very fast. It can be seen in the Yukawa
interactions $y_{1ij}\psi _{iL}^{T}C\chi _{jL}H^{\ast }$ and $y_{2ij}\chi
_{iL}^{T}C\chi _{jL}H$, where $y_k$ are the Yukawa couplings. Omitting the family indices, these
terms leads to the proton decay through interactions like $q_{mL}^{T}y_{1}C%
\overrightarrow{\sigma }\cdot \overrightarrow{\eta }_{m}\ell $ and $\epsilon
_{mnp}q_{mL}^{T}y_{2}C\overrightarrow{\sigma }\cdot \overrightarrow{\eta }%
_{n}q_{pL}$, respectively; here $\epsilon =i\sigma _{2}$ and $m,n,p$ are
color indices, and $\overrightarrow{\eta }_{m}$ is the colored scalar
triplet necessary to unification and which belongs to $(45,1)$-scalar with a relative low mass\footnote{The conditions for proton decay suppression with colored scalar triplet in the $\mathsf{SU(5)}$ context can be found in Refs.~\cite{Dorsner:2009mq,Dorsner:2009cu}.}. The situation is only ameliorated with the introduction of vector-like fermions at unification scale, but without
perfect string unification. It is known that low scales for vector-like fermions can suppress the proton decay in the $\mathsf{SU(5)\times SU(5)}$ theory~\cite{Lee:2016wiy}. As the $\eta $ does not affect too much the evolution of $\alpha _{1}^{-1}$, we can see that its slope is almost constant, as showed in the Fig.~\ref{fig7b}.

In fact,
the $\eta $ was chosen for this reason and, therefore, the couplings have almost the
exact value for perfect string unification. Without vector-like fermions
this occurs near $z=1$, but with a new scale to $\eta$, $M_{2}$, rather above than showed in the
Fig.~\ref{fig7b} and $M_{3}$ somewhat bellow than showed there. We emphasize that any attempt to correct the string unification and keep the proton stable must consider the introduction of only particles with small hypercharge contributions.

In contrast, the SUSY extension is not restricted by proton decay since the
interaction with $(45,1)$ is not necessary. In addition, other
possible interactions, that are allowed in simple $\mathsf{SU(5)}$ context, even in
the minimal SUSY $\mathsf{SU(5)}$\cite{Bajc:2002bv}, are forbidden in the
context of $\mathsf{SU(5)\times SU(5)}$ gauge group, which does not mix heavy
vector-like quarks with light quarks at leading order~\cite%
{Mohapatra:1996fu,Mohapatra:1996iy}. Explicitly, the interactions allowed
due to scalar sector $(\overline{5},1)$ are $q_{mL}^{T}y_{1}^{\prime
}CH_{m}^{c}\ell $ and $\epsilon _{mnp}U_{mL}^{T}y_{2}^{\prime
}CH_{n}^{c}D_{pL}$, where the colored triplet $H_{m}^{c}$ belonging to $(%
\overline{5},1)$ interacts with quarks and leptons doublets of SM, and also interacts with vector-like $\mathsf{SU(2)_L}$
singlet quarks. Choosing reasonable parameters it can leads to proton lifetime
consistent with experimental bounds~\cite{pdg}, similarly to the analysis performed in~\cite{Mohapatra:1996fu,Mohapatra:1996iy}. The supersymmetric model is in this respect phenomenologically safe.

\section{Phenomenological implications and discussion}

\label{sec:pheno}

R-parity is conserved due to the interactions
between the $\mathsf{SU(2)_L}$ doublets of SM be only possible with vector-like $\mathsf{SU(2)_L}$ singlets through the colored scalars. More
specifically, in SUSY GUTs the source of R-parity-breaking is due to
interaction of $\overline{55}10$~\cite{Mohapatra:1996fu,Mohapatra:1996iy,Bajc:2015zja}.
In the class of $\mathsf{SU(5)\times SU(5)}$ models, R-parity-breaking is induced only by $q^{T}\ell D^{c}$ and $%
U^{c}D^{c}D^{c}$, where we are representing all Weyl spinors in the left-handed
fashion, and these interactions do not
lead to R-parity violation involving light fermions.

The interactions with heavy vector-like fermions can
contribute to baryon asymmetry of the universe (BAU) since there is a plenty of CP-sources to this aim, which depends on the specific scalar choices and may generates the asymmetry via the usual leptogenesis mechanism applied to Adjoint $\mathsf{SU(5)}$ schemes~\cite{Blanchet:2008cj}.

Concerning to generalized seesaw mechanism, this model can easily accommodate all low-mass spectrum of fermions with the vector-like fermions at very high scale, both in string model presented here and in the GUT schemes~\cite{EmmanuelCosta:2011wp}. The interaction among vector-like fermions and electroweak fermions in the GUT schemes are keep hold, and it can provide a very elegant explanation to the low-energy scale of SM fermions.

\section{Summary and Conclusions}

\label{sec:conclusions}

We have seen that the string unification, compatible with weakly interacting heterotic string,
is not possible in the context of (low scale) SUSY $\mathsf{SU(5)\times SU(5)}$ symmetry even
at two-loop level. Nevertheless, the high-energy SUSY is still as a possible theory.  Fortunately, the non-SUSY and SUSY $\mathsf{SU(5)\times SU(5)}$ models are possible if the theory is enlarged to incorporate three fermionic generations of adjoint representations of $\mathsf{SU(5)_L}$. However, some problems with proton decay in non-SUSY case can arise due to the introduction of color-triplet, $\eta $, inside of $(45,1)$-scalar representation which is necessary to complete string unification with correct value of $\alpha_{\mathrm{st}}$ at $\Lambda_s$ scale. A more detailed study in this issue would be crucial to determine the proton stability. For both implementations, with SUSY or not, we need very light
triplet fermions $\Sigma_3^F$, which are in the range of LHC experiments.

Besides the difficult of implementation of this solution, no other can emerge easily in the context analyzed
above. The supersymmetric case does not have problems with proton decay and R-parity since the usual R-parity-breaking terms and L-nonconserving terms only interacts SM particles with very heavy vector-like particles, inducing a suppression in both violation contributions. However, $M_S$ is very high, thus excluding theoretically the SUSY schemes from experimental analysis.

Some modifications can occur if we include a different extended patter with many intermediate scales to the unified gauge group $\mathsf{SU(5)\times SU(5)}$. This case is yet more restrictive than the analysis performed here, since new scalar content are inevitable to every breaking of symmetry. They naturally contribute to $f_y$ if they have non-vanishing hypercharge. Another change can be made in a not exact left-right symmetric model, where two distinct gauge couplings for each $\mathsf{SU(5)}$ group are possible. In this case the interpretation of unification must be modified since different couplings means different interactions, thus $\mathsf{SU(5)\times SU(5)}$ is only an intermediated step, and their two different gauge couplings must be unified to some exact superior crossed gauge group, e.g., in $\mathsf{E_{8}\times E_{8}}$ group.

While non-SUSY $\mathsf{SU(5)\times SU(5)}$ may generate problems with proton stability, the SUSY $\mathsf{SU(5)\times SU(5)}$ with $M_S$ at high-scale allows a very small space for string unification due to the hypercharge abundance and it is only possible in the context of Adjoint $\mathsf{SU(5)_{L}}$, which is now an invariant subgroup of $\mathsf{SU(5)\times SU(5)}$. Any new particle beyond the SM of nonvanishing hypercharge can discard these type of models in few years. Only low-energy SM particles are compatible with the string unification and there is must be a desert of new particles until the unification reaches the string scale. Then, we conclude that  besides it is possible to find some scenarios for string unification based on $\mathsf{SU(5)\times SU(5)}$ group, it takes place in very disadvantaged layouts, which can be inferred as a fine-tunning in particle content at intermediated scales. Thus, the string unification in this framework is almost impossible.

\section*{Acknowledgements}

The author thanks H. de Sandes and C. H. Lenzi for useful communications and also thanks the CFTP/IST and the IFGW/Unicamp where this work started and was almost developed. The author is in debit with R. Gonz\'{a}lez Felipe and D. Emannuel-Costa for their
encouragement to write this paper and for some technical discussions. This work was partially supported by the EU project MRTN-CT-2006-035505 and by CNPq through the Grant No. 150416/2011-3.

\appendix

\section{Standard beta coefficients at one- and two-loops}

\label{sec:appendix}
The one-loop $\beta $%
-function coefficients $b_{i}^{z}$, at $M_{Z}$ scale, are given by%
\begin{equation}
b_{i}^{z}=\frac{1}{3}\sum_{R}\left[ s(R)N_{i}(R)\right] -\frac{11}{3}%
C_{2}(G_{i}),
\end{equation}%
with $s(R)$ being definite for the particle representations as
\begin{equation}
s(R)=\left\{
\begin{array}{cl}
1 & \text{complex scalar} \\
2 & \text{chiral fermions} \\
4 & \text{vector-like fermions}%
\end{array}%
\right. .
\end{equation}%
The $C_{2}(G_{i})$ is the Casimir group invariant of adjoint representation
of one Group $G_{i}$\ (equal to $n$ for $\mathsf{SU(n)}$ and null for $\mathsf{U(1)}$). The
functions $N_{i}(R)$ encode the contributions of the group structure as follows,
\begin{equation}
N_{i}(R)=T_{i}(R)\prod_{j\neq i}d_{j}(R),
\end{equation}%
where $d_{j}(R)$ is the dimension of the representation in respect to the
invariant subgroup, $G_{i}$. The Dynkin index is denoted by $T_{i}(R)$ and is settled to be 1/2 for the fundamental representations of $\mathsf{SU(n)}$ groups and $Y^{2}$ for the $\mathsf{U(1)_{Y}}$ Abelian group. Our hypercharge convention sets the hypercharge to coincides with the electrical charges of all singlets, $Y=Q-T_{3L}$.

\subsection{non-SUSY}
For unification of the SM we have
\begin{eqnarray}
b_{y}^{z} &=&\frac{20}{9}N_{g}+\frac{n_{H}}{6}, \\
b_{w}^{z} &=&\frac{4}{3}N_{g}+\frac{n_{H}}{6}-\frac{22}{3}, \\
b_{s}^{z} &=&\frac{4}{3}N_{g}-11,
\end{eqnarray}%
where $N_{g}$ is the number of fermionic generations and $n_{H}$ is the
number of Higgs doublets of $\mathsf{SU(2)_L}$ at $M_Z$ scale. In the SM we have $b_{i}^{z}=\left( 41/6,-19/6,-7\right) $ at one-loop level.

At two-loop level the coefficients $b_{ij}^z$ are given by
\begin{eqnarray}
b_{ij}&=&\left(
\begin{array}{ccc}
0 & 0 & 0 \\
0 & -\frac{136}{3} & 0 \\
0 & 0 & -102%
\end{array}%
\right) +N_{g}\left(
\begin{array}{ccc}
\frac{95}{27} & 1 & \frac{44}{9} \\
\frac{1}{3} & \frac{49}{3} & 4 \\
\frac{11}{18} & \frac{3}{2} & \frac{76}{3}%
\end{array}%
\right) \nonumber\\&+&
n_{H}\left(
\begin{array}{ccc}
\frac{1}{2} & \frac{3}{2} & 0 \\
\frac{1}{2} & \frac{13}{6} & 0 \\
0 & 0 & 0%
\end{array}%
\right),
\end{eqnarray}%
In SM $b_{ij}^{z}=b_{ij}|_{(N_g=3,n_H=1)}$.

\subsection{SUSY}
At one loop level the MSSM with $N_g$ generations and $n_H$ Higgs doublets has the following $\beta$-coefficients\footnote{Here we denote the SUSY $\beta$-coefficients by a $\bar{b}_{\cdots}$.}
\begin{eqnarray}
\overline{b_{y} }&=&\frac{10}{3}N_{g}+\frac{n_{H}}{2}, \\
\overline{b_{w}} &=&2N_{g}+\frac{n_{H}}{2}-6, \\
\overline{b_{s}} &=&2N_{g}-9,
\end{eqnarray}%
and these reads as $\overline{b^{z}}=\left( 11,1,-3\right) $ for the MSSM with $N_g=3$ and $n_H=2$ at $M_S$ scale.

For supersymmetry (in the $\overline{DR}$-scheme at $M_{S}$) the
coefficients of RGEs at two-loop level are given by

\begin{eqnarray}
\overline{b_{ij}}&=&\left(
\begin{array}{ccc}
0 & 0 & 0 \\
0 & -24 & 0 \\
0 & 0 & -54%
\end{array}%
\right) +N_{g}\left(
\begin{array}{ccc}
\frac{190}{27} & 2 & \frac{88}{9} \\
\frac{2}{3} & 14 & 8 \\
\frac{11}{9} & 3 & \frac{68}{3}%
\end{array}%
\right) \nonumber\\
&+& n_{H}\left(
\begin{array}{ccc}
\frac{1}{2} & \frac{3}{2} & 0 \\
\frac{1}{2} & \frac{7}{2} & 0 \\
0 & 0 & 0%
\end{array}%
\right) ,
\end{eqnarray}%

\section{Beta coefficients with new particles at intermediate scale}
\label{sec:appendix_b}

Here we give the beta coefficients for the non standard particle content that have to be
introduced to correct the string unification. We use the $\mathsf{SU(5)_L}$ group notation, given in the $(\mathsf{SU(3)_{c},SU(2)_{L})_{Y}}$ form, as follows:

\begin{equation}
\label{eq:particlesB}
\begin{array}{cc}
\eta =(3,3)_{-1/3}, & U=(3,1)_{2/3}+(\overline{3},1)_{-2/3}, \\
\Sigma _{3}^{S}=\Sigma _{3}^{F}=\Sigma _{3}=(1,3)_{0}, & D=(3,1)_{-1/3}+(%
\overline{3},1)_{1/3}, \\
\Sigma _{8}^{S}=\Sigma _{8}^{F}=\Sigma _{8}=(8,1)_{0}, &
E=(1,1)_{-1}+(1,1)_{1},%
\end{array}%
\end{equation}%
and $F$ refers to Fermions while $S$ refers to Scalars.

\subsection{non-SUSY}

The $\beta$-coefficients at one-loop level are given by
\begin{eqnarray}
\Delta b_{y} &=&\frac{N_{\eta }}{3}+\frac{16}{9}N_{U}+\frac{4}{9}N_{D}+\frac{%
4}{3}N_{E}, \\
\Delta b_{w} &=&2N_{\eta }+\frac{2}{3}N_{\Sigma 3}^{S}+\frac{4}{3}N_{\Sigma
3}^{F}, \\
\Delta b_{s} &=&\frac{N_{\eta }}{2}+2N_{\Sigma 8}^{S}+4N_{\Sigma 8}^{F}+%
\frac{2}{3}N_{U}+\frac{2}{3}N_{D},
\end{eqnarray}
where $N_{i}$ is the number of particles showed in Eq.~\ref{eq:particlesB}.

The $\beta$-coefficients at two-loop level are given by:
\begin{eqnarray}
\Delta b_{ij} &=&\left(
\begin{array}{ccc}
\frac{4}{9} & 8 & \frac{16}{3} \\
\frac{8}{3} & 56 & 32 \\
\frac{2}{3} & 12 & 14%
\end{array}%
\right) N_{\eta }+\left(
\begin{array}{ccc}
0 & 0 & 0 \\
0 & \frac{56}{3} & 0 \\
0 & 0 & 0%
\end{array}%
\right) N_{\Sigma 3}^{S} \nonumber\\
&+&\left(
\begin{array}{ccc}
0 & 0 & 0 \\
0 & 0 & 0 \\
0 & 0 & 42%
\end{array}%
\right) N_{\Sigma 8}^{S}+\left(
\begin{array}{ccc}
0 & 0 & 0 \\
0 & \frac{64}{3} & 0 \\
0 & 0 & 0%
\end{array}%
\right) N_{\Sigma 3}^{F} \nonumber\\
&+&\left(
\begin{array}{ccc}
0 & 0 & 0 \\
0 & 0 & 0 \\
0 & 0 & 48%
\end{array}%
\right) N_{\Sigma 8}^{F}+\left(
\begin{array}{ccc}
\frac{64}{27} & 0 & \frac{64}{9} \\
0 & 0 & 0 \\
\frac{8}{9} & 0 & \frac{38}{3}%
\end{array}%
\right) N_{U} \nonumber\\
&+&\left(
\begin{array}{ccc}
\frac{4}{27} & 0 & \frac{16}{9} \\
0 & 0 & 0 \\
\frac{2}{9} & 0 & \frac{38}{3}%
\end{array}%
\right) N_{D}+\left(
\begin{array}{ccc}
4 & 0 & 0 \\
0 & 0 & 0 \\
0 & 0 & 0%
\end{array}%
\right) N_{E}.
\end{eqnarray}

\subsection{SUSY}

The $\beta$-coefficients at one-loop level are given by:
\begin{eqnarray}
\overline{\Delta b_{y}} &=&N_{\eta }+\frac{8}{3}N_{U}+\frac{2}{3}N_{D}+2N_{E}, \\
\overline{\Delta b_{w}} &=&6N_{\eta }+2N_{\Sigma 3}, \\
\overline{\Delta b_{s}} &=&\frac{3}{2}N_{\eta }+3N_{\Sigma 8}+N_{U}+N_{D}.
\end{eqnarray}

The $\beta$-coefficients at two-loop level are given by:
\begin{eqnarray}
\overline{\Delta b_{ij}} &=&\left(
\begin{array}{ccc}
\frac{4}{9} & 8 & \frac{16}{3} \\
\frac{8}{3} & 72 & 32 \\
\frac{2}{3} & 12 & 17%
\end{array}%
\right) N_{\eta }+\left(
\begin{array}{ccc}
0 & 0 & 0 \\
0 & 24 & 0 \\
0 & 0 & 0%
\end{array}%
\right) N_{\Sigma 3} \nonumber\\
&+&\left(
\begin{array}{ccc}
0 & 0 & 0 \\
0 & 0 & 0 \\
0 & 0 & 54%
\end{array}%
\right) N_{\Sigma 8} +\left(
\begin{array}{ccc}
\frac{128}{27} & 0 & \frac{128}{9} \\
0 & 0 & 0 \\
\frac{16}{9} & 0 & \frac{34}{3}%
\end{array}%
\right) N_{U} \nonumber\\
&+&\left(
\begin{array}{ccc}
\frac{8}{27} & 0 & \frac{32}{9} \\
0 & 0 & 0 \\
\frac{4}{9} & 0 & \frac{34}{3}%
\end{array}%
\right) N_{D}+\left(
\begin{array}{ccc}
8 & 0 & 0 \\
0 & 0 & 0 \\
0 & 0 & 0%
\end{array}%
\right) N_{E}.
\end{eqnarray}

\end{document}